\documentclass[12pt,thmsa]{article}
\usepackage{amssymb}

\usepackage{sw20lart}

\input{tcilatex}
\begin{document}

\author{Jun Luo and Xizhi Zeng \and {\small (Laboratory of Magnetic Resonance and
Atomic and Molecular Physics,} \and {\small Wuhan Institute of Physics and
Mathematics, Chinese Academy of Sciences,} \and {\small Wuhan,Hubei
430071,People's Republic of China)}}
\title{NMR Quantum Computation with a hyperpolarized nuclear spin bulk }
\date{}
\maketitle

\begin{abstract}
We consider two new quantum gate mechanisms based on nuclear spins in
hyperpolarized solid $^{129}Xe$ and HCl mixtures and inorganic
semiconductors. We propose two schemes for implementing a controlled NOT (
CNOT ) gate based on nuclear magnetic resonance ( NMR ) spectroscopy and
magnetic resonance imaging ( MRI ) from hyperpolarized solid $^{129}Xe$ and
HCl mixtures and optically pumped NMR in semiconductors. Such gates might be
built up with particular spins addressable based on MRI techniques and
optical pumping and optical detection techniques. The schemes could be
useful for implementing actual quantum computers in terms of a cellular
automata architecture.
\end{abstract}

\section{Introduction}

The study of quantum computation has attracted considerable attention since
Shor discovered a quantum mechanical algorithm for factorization in
polynomial instead of exponential time \cite{Pwshor} in 1994. In 1996,
Grover \cite{grover} also showed that quantum mechanics can speed up a range
of search applications over an unsorted list of $N$ elements. Hence it
appears that a quantum computer can obtain the result with certainty in $O(%
\sqrt{N})$ instead of $O(N)$ attempts.

In 1982, Benioff \cite{benioff} showed that a computer could in principle
work in a purely quantum-mechanical fashion. In 1982 and 1986, Feynman
showed that a quantum computer might simulate quantum systems \cite{feynman} 
\cite{feynman1}. In 1985 and 1989, Deutsch \cite{deutsch} \cite{deutsch1}
first explicitly studied the question that quantum-mechanical processes
allow new types of information processing. Quantum computers have two
advantages. One is the quantum states can represent a $1$ or a $0,$ or a
superposition of a $1$ and a $0,$which leads to quantum parallelism
computations. The other is quantum computers perform deterministic unitary
transformations on the quantum states.

It is difficult to build up quantum computers for two principal reasons. One
is decoherence of the quantum states. The other is a quantum computer might
be prone to errors, which are troublesomely corrected. The current
developments, however , have showed that the two obstacles might be
surmounted. Shor \cite{shora} \cite{calderbank} and Steane \cite{steane}
discovered that the use of quantum error-correcting codes enables quantum
computer to operate in spite of some degree of decoherence and errors, which
may make quantum computers experimentally realizable. Knill et al. \cite
{knill} also showed that arbitrary accurate quantum computation is possible
provided that the error per operation is below a threshold value. In
addition, it is possible to decrease the influence of decoherences by using
mixed-state ensembles rather than isolated systems in a pure state \cite
{cory} \cite{chuang} \cite{chuang1} .

Among many candidate physical systems envisioned to perform quantum
computations ( such as quantum dots \cite{dot} \cite{dot1} , isolated
nuclear spin \cite{divincenzo}, trapped ions \cite{cz} , optical photons 
\cite{chy}, cavity quantum-electrodynamics \cite{drbh} \cite{tshw} and
nuclear magnetic resonance (NMR) of molecules in a room temperature solution 
\cite{cory} \cite{chuang} etc. ), NMR quantum computers \cite{cory} \cite
{chuang} are particularly attractive because nuclear spins are extremely
well isolated from their environment and readily manipulated with modern NMR
methods. Recently, Chuang and Jones et al . \cite{chuang2} \cite{jones} have
experimentally realized for the first time a significant quantum computing
algorithm using NMR techniques to perform Grover'quantum search algorithm 
\cite{grover}. In addition, NMR quantum computers with two qubits or three
qubits have been used to implement Deutsch's quantum algorithm \cite{chuang3}
\cite{jones1} \cite{lbr}. Cory et al. have experimentally realized for the
first time quantum error correction for phase errors on an NMR quantum
computer \cite{cory98} . However, there are two primary challenges in using
nuclear spins in quantum computers. One is low sensitivity of NMR signals.
The other is that scaling-up to much larger systems with this approach may
be difficult \cite{warren}. Recently, Kane \cite{kane} has presented a
scheme for implementing a quantum computer, using semiconductor physics to
manipulate nuclear spins. This may be an answer to scaling up to produce
useful quantum computers \cite{divincenzo1}. However, the current technology
is difficult to implement Kane's scheme \cite{kane} \cite{divincenzo1}.

Chuang \textit{et al. }first experimentally demonstrated the complete model
of a simple quantum computer by NMR spectroscopy on the small organic
molecule chloroform \cite{chuang2}. Their experimental results showed that
the execution of certain tasks on a quantum computer indeed requires fewer
steps than on a classical computer. However, building a practical quantum
computer by the use of NMR techniques poses a formidable challenge. Steps of
circumventing these problems based on the bulk spin resonance approach to
build quantum computers can include increasing the sample size, using
coherence transfer to and from electrons, and optical pumping to cool the
spin system \cite{chuang4}.

DiVincenzo \textit{et al.} have suggested that a ''solid state'' approach to
quantum computation with a $10^{6}$ qubit might be possible \cite{dot1}.
Solid-state devices open up the possibility of actual quantum computers
having realistic applications \cite{divincenzo1}. In the paper, we shall
discuss the implementation of a controlled NOT gate between two spins with a
spin-hyperpolarized bulk. It should be noted that an important progress in
optical pumping in solid state nuclear magnetic resonance (OPNMR) has been
made \cite{tr}. Recently, interest has been growing in exploitation of
optical pumping of nuclear spin polarizations as a means of enhancing and
localizing NMR signals in solid state nuclear magnetic resonance \cite{tr}.
The principal work has been concentrated in the following two areas. The
polarization of $^{129}Xe$ can be enormously enhanced through spin-exchange
with optically pumped alkali-metal vapor \cite{happer}. The NMR signals from 
$^{129}Xe$ nuclei have enhanced to $\thicksim 10^{5}$ times the thermal
equilibrium value \cite{happer}. The spin-lattice relaxation of the $%
^{129}Xe $ nuclei in the polarized solid $^{129}Xe$ is exceptionally slow
(such as relaxation times longer than 500h \cite{happer1}). In addition, the
experimental results \cite{happer1} \cite{pines} \cite{happer2} \cite{pines1}
\cite{pines2} have shown that laser-polarized $^{129}Xe$ nuclei can be used
to polarize other nuclei that are present in the lattice or on a surface
through cross relaxation or cross polarization. Optical pumping NMR (OPNMR)
techniques of inorganic semiconductors (such as GaAs, and InP \textit{et al.}%
) \cite{tr} at low temperatures can enhance NMR signals in solids, which can
circumvent the two problems in solid state NMR, that is, its relative low
sensitivity and its lack of spatial selectivity. In OPNMR, spatial
localization of the NMR signals can be achieved through spatial localization
of the optical absorption and through cross polarization or relaxation
mechanisms \cite{tr} \cite{gbskp} \cite{ty}. In this paper, we shall propose
two schemes for implementing a controlled NOT (CNOT) gate in quantum
computers based on NMR spectroscopy and magnetic resonance imaging from
hyperpolarized solid $^{129}Xe$ and HCl mixtures and OPNMR in the solid
state nuclear magnetic resonance of inorganic semiconductors ( quantum well
and quantum dot ).

The paper is organized as follows. In Sec.II, we introduce the model for the
controlled NOT gates in terms of hyperpolarized solid $^{129}Xe$ and HCl
mixtures. In Sec.III we present a scheme for implementing a CNOT gate based
on OPNMR in inorganic semiconductors. In Sec.IV, we present implications for
experiments on our schemes.

\section{Quantum computation with hyperpolarized $^{129}Xe$ and HCl solid
mixtures}

To realize quantum computation, it is necessary to have the nonlinear
interactions in a system. These nonlinear interactions can simultaneously be
influenced externally in order to control states of the system. Meanwhile,
it is required that the system can be extremely well isolated from its
environment so that the quantum coherence in computing is not rapidly lost.
In a solid, there are dipolar couplings between two spin systems, which
result in broader NMR lines and cross relaxation among spin systems. The
interactions in solids can be controlled with complex radio-frequency (rf)
pulse sequences (such as decoupling pulse sequences). In general, the
interactions in solids are so strong that the eigenstates are not the simple
spin product states and logical manipulation are more complex \cite{warren}.
However, for special solids (such as $^{129}Xe$ and $^{1}HCl$ , and $%
^{129}Xe $ and $^{13}CO_{2}$ mixtures), since the homonuclear spin system is
diluted by other spin system , the spin dipolar interactions between two
homonuclei are weak and the solid mixtures \hspace{0.05in}may be homogeneous
so that the solids could have a resolved dipolar structure. A full quantum
mechanical treatment of the spin system is in order \cite{ernst}.

Spin-exchange optical pumping can produce hyperpolarized $^{129}Xe$ (with an
enhanced factor of about $10^{5}$ )\cite{happer}. The hyperpolarized $%
^{129}Xe$ gas can be frozen into a hyperpolarized $^{129}Xe$ solid with
little or no loss of $^{129}Xe$ nuclear spin polarization $\cite{happer1}$.
It should be noted that nuclear spin polarization of $^{129}Xe$ , which is
produced with spin-exchange optical pumping, does not depend on the strength
of magnetic fields. Therefore, the NMR experiments can be performed in low
fields produced by the general electromagnets or the magneto irons. NMR
signals with sufficient signal-to-noise ratio from a hyperpolarized $%
^{129}Xe $ solid are available on a single acquisition. At low temperatures,
the spin-lattice relaxation of the $^{129}Xe$ nuclei in the hyperpolarized
solids is extremely slow. For example, the $^{129}Xe$ spin polarization
lifetime $T_{1}$ is hundreds of hours at 1 KG below 20K \cite{happer1}.
Linewidth of NMR signals from a hyperpolarized $^{129}Xe$ solid is tens of
Hz \cite{zeng}. This indicates that the $^{129}Xe$ nuclei in a
hyperpolarized solid can be relatively well isolated from their environment.
Through dipolar-dipolar interactions, enhanced nuclear spin polarization of $%
^{129}Xe$ can be transferred to other nuclei ($^{1}H$ and $^{13}C$ $etc.$)
on a surface \cite{pines1} \cite{happer2}, in a solid lattice \cite{pines} 
\cite{happer1}and a solution \cite{pines2}. The experimental results have
indirectly shown that the sign of the $^{129}Xe$ polarization can be
controlled by the helicity of the pumping laser or the orientation of the
magnetic field in the optical pumping stage \cite{pines} \cite{pines2}. It
is interesting to note that the signs of polarization of other nuclei ($%
^{1}H $ and $^{13}C$) depend on those of polarization of $^{129}Xe$ nuclei
in the cross polarization experiments \cite{pines} \cite{pines2}. The signs
of polarization of other nuclei ($^{1}H$ and $^{13}C$) are the same as $%
^{129}Xe $ nuclei \cite{pines} \cite{pines2}.

In quantum computation, Barenco \textit{et al.} \cite{ba} have shown that
single-spin rotations and the two-qubit ''controlled''-NOT gates can be
built up into quantum logic gates having any logical functions. In the
following, we shall show how to build up a ''controlled''NOT gate based on
NMR signals from a hyperpolarized $^{129}Xe$ solid.

Before our discussions, we first show how to prepare a hyperpolarized $%
^{129}Xe$ and HCl solid mixture. First, spin-exchange optical pumping
produces hyperpolarized $^{129}Xe$ gas at a 25G magnetic field. Secondly,
the polarized xenon is mixed with 760 Torr of HCl at room temperature and
the mixture is rapidly frozen into the sample tube in liquid $N_{2}$.

In solids, the dominant mechanism of spin-spin relaxation is the
dipole-dipole interaction. The experimental results \cite{pines} \cite
{pines2} \cite{happer2} have shown that the transfer of a large nuclear
polarization from $^{129}Xe$ to $^{1}H$ (or $^{13}C$) can be controlled by
cross polarization techniques in the solids. Hyperpolarized $^{129}Xe$ ice
can be used to polarize $^{131}Xe$ \cite{happer1} and $^{13}C$ ($CO_{2})$ 
\cite{pines} trapped in the xenon lattice through thermal mixture in low
fields. Pines \textit{et al. } \cite{pines1} have shown that high-field
cross polarization methods can make magnetization transfer between two
heteronuclear spin systems selective and sensitive.

In a quantum computer, logic functions are essentially classical, only
quantum bits are of quantum characteristic (quantum superpositions) \cite
{grover}. In the $^{129}Xe$ and HCl solid mixture, one can use complex pulse
sequences and dipole-dipole interactions between $^{129}Xe$ and $^{1}H$ to
manipulate and control two qubits ( $^{129}Xe$ and $^{1}H$ ). In the
following, we shall discuss the scheme for implementing a controlled NOT
(CNOT) gate in quantum computers based on NMR spectroscopy and magnetic
resonance imaging from the hyperpolarized solid $^{129}Xe$ and HCl mixtures.

Since $^{129}Xe$ and $^{1}H$ in the solid are a weakly- coupled two-spin IS
system with a resolved structure, in the doubly rotating frame, rotating at
the frequencies of the two applied r.f. fields, the Hamiltonian of this
system can be written as \cite{ernst}

$\mathbf{H=}\Omega _{I}I_{Z}+\Omega _{S}S_{z}+2\pi J_{IS}I_{z}S_{z}+\omega
_{1I}I_{x}+\omega _{1S}S_{x}$

where $\Omega _{I}$ and $\Omega _{S}$ are the resonance offsets, and $\omega
_{1I}$ and $\omega _{1S}$ are the two r.f. field strengths, respectively. $%
\omega _{1I}$ and $\omega _{1S}$ can be used to perform arbitrary
single-spin rotations to each of the two spins (I and S) with selective
pulses \cite{ernst}.

In the above Hamiltonian, the Hamiltonian $\mathbf{H}_{IS}=2\pi
J_{IS}I_{z}S_{z}$ leads to the following evolution operator

$\widehat{R}_{zIS}(J_{IS}\tau \pi )=e^{i2\pi J_{IS}\tau I_{z}S_{z}}$

$\qquad =\cos (J_{IS}\tau \pi /2)+i\sin (J_{IS}\tau \pi /2)\left| 
\begin{array}{cccc}
1 & 0 & 0 & 0 \\ 
0 & -1 & 0 & 0 \\ 
0 & 0 & -1 & 0 \\ 
0 & 0 & 0 & 1
\end{array}
\right| $.

when $\tau =\frac{1}{2J_{IS}},\widehat{R}_{zIS}(J_{IS}\tau \pi )=\frac{\sqrt{%
2}}{2}\left| 
\begin{array}{cccc}
1+i & 0 & 0 & 0 \\ 
0 & 1-i & 0 & 0 \\ 
0 & 0 & 1-i & 0 \\ 
0 & 0 & 0 & 1+i
\end{array}
\right| $.

It is easy to perform single- spin rotations with arbitrary phase $\phi $
with the modern pulse NMR techniques \cite{ernst}. For example, one can
perform single-spin rotations via composite z-pulses \cite{freeman} and free
precession \cite{jones1} .

In the ideal case, the operator to perform CNOT gates can be written as

$\widehat{C}_{CNOT}=\left| 
\begin{array}{cccc}
1 & 0 & 0 & 0 \\ 
0 & 1 & 0 & 0 \\ 
0 & 0 & 0 & 1 \\ 
0 & 0 & 1 & 0
\end{array}
\right| $.

A ''controlled ''NOT gate can be realized by the following pulse sequences 
\cite{chuang} \cite{chuang4}

$\widehat{C}_{1AB}=\widehat{R}_{yA}(-\pi /2)\widehat{R}_{zB}(-\pi /2)%
\widehat{R}_{zA}(-\pi /2)\widehat{R}_{zAB}(\pi /2)\widehat{R}_{yA}(\pi /2)$

$\;\;\;\;\;\;=\sqrt{-i}\left| 
\begin{array}{llll}
1 & 0 & 0 & 0 \\ 
0 & 1 & 0 & 0 \\ 
0 & 0 & 0 & 1 \\ 
0 & 0 & 1 & 0
\end{array}
\right| \;\;\;\;\;\;(1),$

or

$\widehat{C}_{2AB}=\widehat{R}_{yB}(\pi /2)\widehat{R}_{zAB}(\pi /2)\widehat{%
R}_{xB}(\pi /2)$

$\;\;\;\;\;=\left| 
\begin{array}{llll}
(-1)^{1/4} & \ \;0 & \;0 & \;0 \\ 
\;0 & -(-1)^{3/4} & \;0 & \;0 \\ 
\;0 & \;\,0 & \;0 & (-1)^{1/4} \\ 
\;0 & \;\,0 & (-1)^{3/4} & \;0
\end{array}
\right| \;\;\;(2),$

where A and B represent ''target''qubit ( $^{1}H$ (HCl)) and
''control''qubit ( $^{129}Xe$), respectively. In eq. (1), one can not only
apply composite z-pulses \cite{freeman} actively to perform $\widehat{R}%
_{zA}(-\pi /2)$ and $\widehat{R}_{zB}(-\pi /2)$ but also let a two-spin AB
system freely evolve in periods of precession under Zeeman Hamiltonians.
When using pulse-sequences in Eqs. (1) and (2), one can obtain a similar
effect on quantum computation \cite{chuang4} . This involves an important
problem, that is, optimized rf pulse sequences. It is important to eliminate
unnecessary pulses.

The CNOT operations can also be performed on the basis of the
cross-polarization experiments in solid state NMR. Under some experimental
conditions, one uses pumping laser with helicities $\sigma ^{-}$ and $\sigma
^{+}$ to produce positive or negative polarizations of $^{129}Xe$ through
spin-exchange laser pumping techniques before one performs the CNOT
operations. If polarization of $^{129}Xe$ nuclei is negative, no operation
is needed. If polarizations of $^{129}Xe$ nuclei and $^{1}H$ nuclei are
positive, one can use the cross-polarization pulse sequences shown in Fig.1a
to perform the controlled rotation operations. If polarizations of $^{129}Xe$
nuclei and $^{1}H$ nuclei are respectively positive and negative, one can
use the pulse sequences in Fig.1b to perform the CNOT gate operations.

In order to yield a large sensitivity enhanced for $^{1}H$, before
performing CNOT gates, we perform cross polarization experiments of the $%
^{1}H$ and $^{129}Xe$ system. In addition, in order to enhance sensitivity
of NMR signals, one can selectively consider a hyperpolarized $^{129}Xe$ and
a $^{1}H$ as the ``target'' qubit and the ``control'' qubit respectively. If
we use three types of gases ( $^{129}Xe$ , $^{1}HCl$ and $^{13}CO_{2}$) to
prepare the NMR sample, we can perform quantum logic gates with three qubit (%
$^{129}Xe$ , $^{1}H$ and $^{13}C$). If we respectively use different
pressures of $^{129}Xe$ , $^{1}HCl$ and $^{13}CO_{2}$, we can increase or
decrease the homonuclear dipole-dipole interaction. That is because when the
relative pressure of the other gases is increased or decreased, the distance
r between two homonuclei is increased or decreased.

As we know, the principal limitation of solid state NMR is its relatively
low sensitivity. In addition, another limitation of solid state NMR is its
lack of spatial selectivity and extremely broad NMR lines. This is because
the NMR signals from an inhomogeneous sample are bulk-averaged \cite{tr} and
the dipole-dipole interactions between two spins are strong. Therefore, it
is difficult to assign particular logic gates to regions within the bulk
sample. Constructing the actual quantum circuit based on NMR could be an
important issue. Now, the actual quantum circuit based on liquid state NMR
is constructed only \textit{in time} by using different pulse sequences. Can
we build up actual quantum circuits both \textit{in time }and \textit{in
space}? It is interesting to note that NMR signals from $^{129}Xe$ ice have
relatively high sensitivity and that NMR lines are relatively narrow \cite
{zeng}. It is possible to construct the actual quantum circuit based on
hyperpolarized $^{129}Xe$ ice \textit{in space }using magnetic resonance
imaging (MRI) and NMR spectroscopy techniques. If one uses gradient fields 
\cite{cory}, one can make the quantum logic gates localized in space in a
bulk sample. Therefore logic gates having any logical functions can be
performed not only\textit{\ in space} but also \textit{in time. }%
Entanglement between spins of two different cells may be realized by
homonuclear spin diffusion.

\section{Quantum computation based on optically pumped NMR of semiconductors}

Recently, great progress in optically pumped NMR (OPNMR) in semiconductors,
single quantum dot, as well as quantum wells has been made \cite{tr} \cite
{gbskp} \cite{ty}. An optical pumping technique can be used to enhance and
localize nuclear magnetic resonance signals. This method has greatly
improved spatial resolution and sensitivity of NMR signals.

Spatial localization of the NMR signals from solids can be achieved through
spatial localization of the optical absorption and through subsequent
manipulations and transfers of the optically pumped nuclear spin
polarization \cite{tr}.

Enhanced NMR signals can be detected either indirectly by optical techniques 
\cite{tr} or directly by conventional rf pick up coils \cite{tr} . Optical
detection can extremely sensitively measure NMR signals from a single
quantum dot ( $\leq 10^{4}$ nuclei ) \cite{gbskp}. Here we propose a scheme
for the implementation of quantum logic gates by using OPNMR in solids,
where the electron spin $\overrightarrow{S}$ and the nuclear spin $%
\overrightarrow{I}$ are respectively considered as the ``control''qubit and
the ``target''qubit.

For semiconductors with zinc blende structures ( such as, Si and GaAs 
\textit{etc. ), }spin polarization of conduction electrons in semiconductors
can be produced by near-infrared laser light with circularly polarized light
working at the band gap \cite{tr}. Since electron and nuclear spins are
coupled by the hyperfine interaction, polarization is transferred between
electrons and nuclei by the spin flip-flop transitions. Through many optical
pumping cycles, large polarization of nuclear spins can be achieved at low
temperature. When the large nuclear spin polarizations are produced, one can
directly detect NMR signals from hyperpolarized nuclei in the semiconductors
with conventional NMR techniques or optical techniques.

In addition, since optical detection of NMR is extremely sensitive, which
can detect NMR signals from fewer than $10^{4}$ \cite{gbskp} nuclei, the
optical methods are naturally preferable to the conventional NMR methods.
Optical detection is mainly based on the following two mechanisms. In
direct-gap semiconductors, a conduction electron can emit a circularly
polarized photon through recombining with a hole in the hole band. The
degree of circular polarization of the photoluminescence depends on the
polarization of the conduction electrons. The hyperfine interaction leads to
spin polarization transfer between electrons and nuclei. Therefore it is
possible to detect NMR in direct-gap semiconductors indirectly by optical
detection methods \cite{tr}. In addition, under NMR conditions,
hyperpolarized nuclei in semiconductors can act back on electron spins by
the hyperfine interaction so that they shift electron Zeeman levels (
Overhauser shift ) \cite{gbskp} and change polarization of electron spins.
This is because hyperpolarized nuclei exert a magnetic field ( called the
nuclear hyperfine field ) on the electron spins \cite{gbskp}. The nuclear
hyperfine field is directly proportional to the nuclear spin polarization.
Radio-frequency pulse near an NMR transition can change the strength and
direction of this field. The net magnetic field felt by the electron spins
depends on the combined action of this field and the externally applied
magnetic field. The shift of the conduction electron Zeeman levels results
from the change of the net magnetic field. It is possible to measure this
Overhauser shift \cite{gbskp} and the polarization of the photoluminescence
through sensitive optical spectroscopies with tunable lasers and highly
sensitive detectors under NMR conditions \cite{gbskp}. Therefore one
indirectly measures NMR in the direct-gap semiconductors by optical
detection methods.

In the following, we shall discuss how to prepare and measure the states of
the nuclear spins in the direct-gap semiconductors and how to perform the
controlled rotation operation on the nuclear spins in our scheme.

For simplicity, we shall take the direct-gap semiconductor InP into account,
which has electronic levels similar to GaAs ( see Fig. 2a )\cite{tr} . Near
the $\Gamma $ point ( electronic Bloch wavevector k is equal to zero ), an
excess of conduction electrons with $m_{1/2}=+1/2$ can be produced by using
circularly polarized light with helicity $\sigma ^{-}$ tuned to the band gap
( E$_{g}\approx 1.42eV$ in InP near 0 K ) \cite{tr} . The direction of the
conduction electron spin polarization can be controlled by circularly
polarized light with different helicities ( such as $\sigma ^{-}$ or $\sigma
^{+}$ ). Electron spin polarization can be transferred to the $^{31}P$
nucleus by the hyperfine interaction. In addition, the $^{31}P$ nuclear
resonance frequency change is directly proportional to the electron spin
polarization $\langle S_{z}\rangle $ \cite{ty}, i.e. $\Delta f=A\rho
(z^{^{\prime }})\langle S_{z}\rangle $ \cite{ty}, where A is a coupling
constant, $\rho (z^{^{\prime }})$ is the conduction electron density envelop
function, $z^{^{\prime }}$ is the displacement of nucleus from a conduction
electron and $\langle S_{z}\rangle $ is electron spin polarization. The
nuclear resonance frequency of $^{31}P$ can be written as: $\nu =\gamma
B_{0}/2\pi +\Delta f,$ where $\gamma $ is the gyromagnetic ratio and $B_{0}$
is the externally applied magnetic field. Therefore one can use the observed
NMR frequency shift based on positive or negative electron spin polarization 
$\langle S_{z}\rangle $ to control $^{31}P$ nuclear rotations by using
selective r.f. pulses. For example, only when the electron spin polarization
is positive, a radio pulse required to flip spin can be used to selectively
change the states of the nucleus. When the electron spin polarization is
negative, one has no use for doing anything. Therefore, one can perform the
controlled rotation operations of $^{31}P$ nuclear spin on the basis of
different directions of electron spin polarizations.

The purpose of optical pumping in the computer is to control the hyperfine
interaction between electrons and nuclei, indirectly to mediate nuclear spin
interactions, to produce electron and nuclear spin polarization and
indirectly to measure nuclear spin polarization. For example, when the
valence band electrons are excited to the conduction band near the $\Gamma $
point, the large hyperfine interaction energy is yielded. This is because
near the $\Gamma $ point, the conduction band is primarily composed of
wavefunctions with $s$ orbits so that the electron wavefunctions are
concentrated at the nucleus. When one uses a laser to pump a cell between
two cells in the semiconductors (see Fig. 2b ), it can enhance nuclear
dipole-dipole interactions and mediate the indirect nuclear spin coupling.
This is because when the electrons are excited to the conduction band by a
laser, the conduction electron wavefunction extends over large distance
through the crystal lattice and large electron spin polarizations are
produced. Electron spin polarization can be transferred to nuclear spins by
the hyperfine interaction. As soon as larger nuclear spin polarizations are
produced, the nuclear spins act back on the electron and shift the electron
Zeeman energy ( Overhauser shift ) \cite{gbskp}. By measuring the magnitude
of the Overhauser shift under NMR conditions, it is possible to measure the
states of nuclei with the Raman spectroscopy \cite{gbskp}.

On the basis of the above discussions, in the following, we shall discuss
how to prepare the cell magnetization in an initial state, how to perform
the controlled rotation operations of $^{31}P$ nuclear spin, how to couple
the effective pure states of the two adjacent cells and how to measure the
effective states of logic gates at different cells.

Optical pumping in the computer can be used to prepare the electron spin
states and the nuclear spin states, i.e. control qubits and target qubits.
As we have seen, one can pump the electrons in valence band into positive or
negative polarizations of conduction electrons by using circularly light
with different helicities ( $\sigma ^{-}$ or $\sigma ^{+}$ ). Therefore an
initial state in a cell can be loaded with circularly polarized light with
different helicities ( $\sigma ^{+}$ or $\sigma ^{-}$ ). For example, in
order to prepare electron and nuclear spins up ( at a magnetic field ), spin
up of the conduction electrons can be produced with circularly polarized
light with helicity of $\sigma ^{-}$ tuned to the band gap, and spin up of $%
^{31}P$ nuclei can also be prepared by the hyperfine interaction. Similarly,
one can prepare electron and nuclear spins down with circularly polarized
light with helicity of $\sigma ^{+}$ . In Fig.2b, one uses many laser beams
with different helicities ( $\sigma ^{+}$ or $\sigma ^{-}$ ) to initial
logic gates at different cells in the different states.

The controlled rotation operations of $^{31}P$ nuclear spins can be
performed by r.f pulses with different frequencies on the basis of the
states ( spins up or down ) of the conduction electrons at different cells,
as has been discussed above.

In Fig.2b, entanglements of effective pure states between cell 1 and cell 3,
might be performed by the following means. Between two logic gates ( such as
1 and 3 in Fig.2b ), if one uses laser with higher power to pump cell 2, one
can produce much more conduction electrons near cell 2 and obtain higher
polarization of the conduction electrons at cell 2, laser pumping at cell 2
results in the electron wavefunction at the cell extending over large
distances through the crystal lattice. The electron spin dipole-dipole
interactions between these two gates ( 1 and 3 ) are increased through the
conduction electrons of the 2 cell mediating electron spin dipole-dipole
interactions between cell 2 and cell 1, and cell 2 and cell 3. Therefore one
can use pumping laser light indirectly to mediate an indirect coupling
between $^{31}P$ qubits of two cells by means of the $^{31}P$
electron-nuclear hyperfine interaction. If one uses pumping laser with low
or no power to pump cell 2, one can decrease the indirect interactions
between $^{31}P$ qubits of two cells so that two gates 1 and 3 can
independently work. As the distance between two cells is increased, the
interactions between distant logic spin will no longer be effective,
Fortunately, universal quantum computation is still possible with just local
interactions \cite{chuang4} \cite{lloyd}. This is because one can use a
cellular automata architecture to perform any function computations in a
linearly-increasing computational time with system size due to massage
passing \cite{lloyd} .

The $^{31}P$ nuclear spin states at a cell can indirectly be measured with
the Raman spectroscopy \cite{gbskp}. This is because when $^{31}P$ nuclear
spin is up or down, the magnitude of the Overhauser shift under NMR
conditions is different \cite{gbskp}.

Can we construct different logic gates in space in semiconductor? i.e. can
one make logic gates addressable? In Fig.2b, similar to the scheme of
trapped ion quantum computer \cite{cz} , we use N\ different laser beams
with different helicities to pump and detect different cells in space in the
InP semiconductor, together with the externally applied gradient magnetic
fields and r.f. gradient pulses, so that we can build up any logic gates in
space.

\section{Conclusion}

We have described the two schemes for implementing the controlled NOT gates
in NMR quantum computers. It is possible to realize these two proposals with
current techniques. Optical pumping in solid state NMR can circumvent the
two problems of relatively low sensitivity and lack of spatial selectivity
in solid state NMR. Therefore it is possible to construct quantum logic
gates in space. The schemes could be useful for implementing actual quantum
computers in terms of a cellular automata architecture. It should be noted
that nuclear spin polarization in optical pumping solid state NMR does not
depend on the strength of magnetic fields. The experiments can be performed
in low fields produced by the general electromagnets or the magneto irons.
The experimental demonstration of our proposal with optical pumping in solid
state NMR and modern NMR techniques is possible.

\medskip

{\Large Acknowledgments}

This work has been supported by the National Natural Science Foundation of
China.

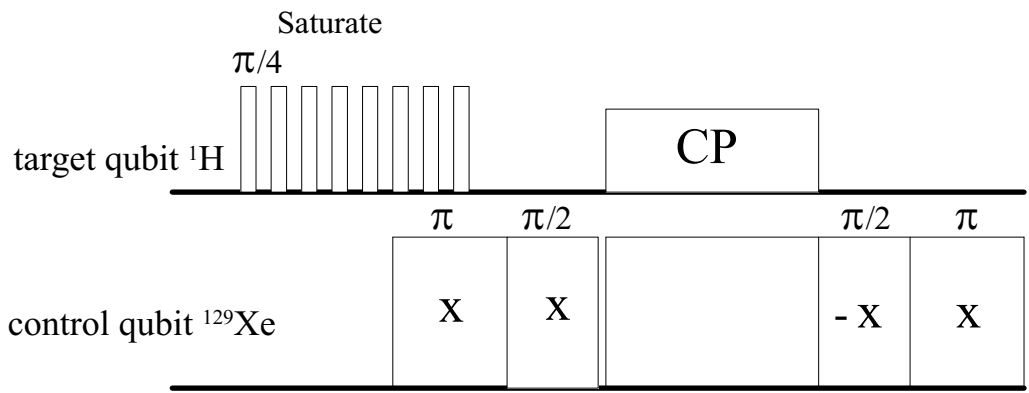
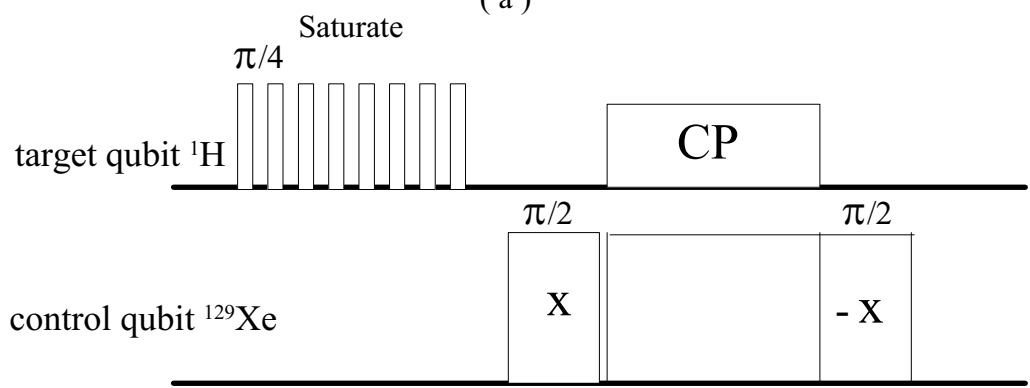
\end{document}